\documentclass[aps,preprint]{revtex4}%
\usepackage{amsfonts}
\usepackage{amsmath}
\usepackage{amssymb}
\usepackage{graphicx}%
\providecommand{\U}[1]{\protect\rule{.1in}{.1in}}

\begin{document}
\title{Semiclassical expansion of the Slater sum for position dependent mass
distributions in $d$ dimensions}
\author{K. Berkane and K. Bencheikh}
\affiliation{Laboratoire de physique quantique et syst\`{e}mes dynamiques. Facult\'{e} des
sciences. D\'{e}partement de physique. Universit\'{e} Ferhat Abbas Setif,
S\'{e}tif 19000, Algeria}
\date{\today}

\begin{abstract}
\baselineskip=17pt

We consider hamiltonian \textbf{systems} with spatially varying effective mass
and slowly varying local potential in $d$ dimensions. The Slater sum is
defined as the diagonal element of the Bloch propagator. We derive a gradient
expansion of the Slater sum up to the second order. We will show that the
derived analytical expression is valid for $d=1,2,3$ and$\ 4$. A numerical
example is shown to highlight the effect of the spatially varying effective
mass\textbf{.}

\end{abstract}
\maketitle

\address{ $\ \ \ \ \ \ \ \ \ \ \ \ \ \ (1)$ D\'{e}partement de Physique.
Laboratoire de Physique Quantique et\\
\ \ \ \ \ \ \ \ \ \ \ \ \ \ \ \  \ \ \ Syst\`{e}mes Dynamiques. Universit\'{e} de S\'{e}tif, S\'{e}tif 19000, Algeria.\\
$\ \ \ \ \ \ \ \ \ \ \ \ \ \ \ (2)$ Department of Physics and Astronomy
University of Missouri,\\
\ \ \ \ \ \ \ \ \ \ \ \ \ \ \ \ \ \ \ \ \ Columbia, Missouri 65211, USA.}

\address{ $\ \ \ \ \ \ \ \ \ \ \ \ \ \ (1)$ D\'{e}partement de Physique.
Universit\'{e} de s\'{e}tif.\\
\ \ \ \ \ \ \ \ \ \ \ \ \ \ Setif 19000, Algeria\\
\ \ \ \ \ \ \ \ \ \ \ \ \ \  $(2)$ Department of Physics and Astronomy
University of Missouri,\\
\ \ \ \ \ \ \ \ \ \ \ \ \ \ Columbia, Missouri 65211, USA}

\address{$(1)$ D\'{e}partement de Physique. Universit\'{e} de s\'{e}tif.\\
Setif 19000, Algeria\\
$(2)$ Department of Physics and Astronomy University of Missouri, columbia,\\
Missouri 65211, USA}


\newpage

\baselineskip=17pt

\section{Introduction}

Quantum mechanical system with a spatially varying effective mass have
attracted a lot of attention and inspired intense research activities during
recent years. Special applications are carried out in the study of electronic
properties of semiconductors \cite{Bastard}, quantum wells and quantum dots
\cite{Serra97}, \cite{Khordad}, 3He clusters \cite{Gora}, quantum liquids
\cite{Arias} graded alloys and semiconductor heterostructures \cite{Weisbuch}
\ldots etc. These studies stimulated a lot of work in the literature on the
development of methods and techniques for studying systems with mass that
depends on position.\ Moreover, the Bloch propagator or its Fourier transform
namely the Green function are of prime significance since they contain all
quantum mechanical informations on the system. To our knowledge the existing
list devoted to the study on the Bloch propagator of quantum systems involving
position dependent effective mass is very short and is limited to the one
dimensional case \cite{chetouani2009}, \cite{alhaidari}. This has motivated
the present work. We wish to obtain approximate analytical expression for the
propagator or at least for its diagonal parts in spatial dimension $d$. The
latter is called the Slater sum. For this purpose, we make use of a
semiclassical approximation to derive analytical expressions of the Slater sum
up to order $\hbar^{2}$ when the effective mass is allowed to be position
dependent. For any exact result of the Slater sum, when expanded up to order
$\hbar^{2}$, our resulting expression may constitute a good test since it is
valid for arbitrary potential and spatially varying effective mass.

Semiclassical $\hbar$ expansion may be generated through a variety of
procedures in order to obtain\textbf{ }the $\hbar$ expansion of the density
matrix. We mention, for instance,\ the partition function method of
Wigner-Kirkwood and further development by Bhaduri and collaborators (see Ref.
\cite{Brack_Bhaduri2003} and references cited therein), the Kirzhnits
expansion using commutator formalism (see Ref. \cite{Dreizler 1990} and
references cited therein), \cite{salasnich}, \cite{Rasanen_2012} and the
purely algebraic method introduced by Baraff and Borowich
\cite{Braff_Borowitz} and developed by Grammaticos and Voros
\cite{Grammaticos_79}, based on the Wigner transform of operators. The latter
method is particularly suitable for position dependent mass Hamiltonians
\cite{Bartel}. These authors derived the semiclassical $\hbar$ expansions for
the 3-dimensional one particle density and also for other densities of
physical interest when the kinetic energy operator of the one-particle
Hamiltonian contains a spatially dependent effective mass. Later on, we have
generalized \cite{bencheikh_Berkane2004}, \cite{Berk_bench2005} such
expansion, up to order $\hbar^{2}$, for systems with effective mass
distribution and reduced dimensionality, i.e, $d=1,2$ dimensions. It should be
noted that corrections of order $\hbar^{2}$ generate second order gradient
corrections not only in the one-body potential but also in the effective mass
distribution. Here, we are interested in obtaining the gradient expansion
of\textbf{ }the Slater sum in $d=1,2$ and $4$ spatial dimensions for
hamiltonians with position dependent mass (the result in $d=3$ is already
known ).

The the paper is organized as follows. In Section $2$ we briefly recall some
basic definitions concerning the use of the Bloch propagator\ and its
relationship to the density matrix. Starting from the semiclassical $\hbar$
expansion for the particle density, we derive, in Section $3$, the
corresponding expansion up to second order in $\hbar$ for the\ Slater sum in
$d=1,2$ and $4$ \ spatial dimensions for hamiltonians with position dependent
mass. A general analytical expression is found in terms of the space dimension
$d$ . Section $4$ provides an illustrative numerical example. Finally, a
conclusion is given in Section $5.$

\section{Basic Concepts}

Consider a system of $N$ \ noninteracting fermions with spatially varying
effective mass $m^{\ast}(\overrightarrow{r})$ moving in a smooth potential
$U(\overrightarrow{r})$. Throughout the present study, we shall be working
with the one-body Hamiltonian given by%

\begin{equation}
H=-\frac{\hbar^{2}}{2m_{0}}\overrightarrow{\nabla}.f(\overrightarrow
{r})\overrightarrow{\nabla}+U(\overrightarrow{r}) \tag{1}%
\end{equation}
where $f(\overrightarrow{r})=m_{0}/m^{\ast}(\overrightarrow{r})$ denotes the
ratio of the free particle mass $m_{0}$ to the position dependent effective
mass $m^{\ast}(\overrightarrow{r})$ and where we use, as done in the majority
of work on the subject, the symmetric ordering form of mass and momentum in
the kinetic energy term of $H.$

Let $\varphi_{n}(\overrightarrow{r})\ $be the eigenfunctions of $H$ and
$\varepsilon_{n}$ the corresponding eigenvalues, i.e; $H\varphi_{n}%
(\overrightarrow{r})=\varepsilon_{n}\varphi_{n}(\overrightarrow{r})$. At zero
temperature, the single-particle density matrix of the system $\rho
(\overrightarrow{r},\overrightarrow{r}^{,})$ is given by
\begin{equation}
\rho(\overrightarrow{r},\overrightarrow{r}^{,})=%
{\displaystyle\sum\limits_{n}}
\varphi_{n}^{\ast}(\overrightarrow{r})\varphi_{n}(\overrightarrow{r}%
^{,})\Theta(\lambda-\varepsilon_{n}) \tag{2}%
\end{equation}
where $\lambda$ is the Fermi energy and $\Theta(x)$ is the Heaviside step
function which allows to restrict the sum over occupied states only.

Given the above density matrix, the Bloch propagator $C(\overrightarrow
{r},\overrightarrow{r}^{,};\beta)$, defined as [see for instance
\cite{Brack_Bhaduri2003}]
\begin{equation}
C(\overrightarrow{r},\overrightarrow{r}^{,};\beta):=\left\langle
\overrightarrow{r}\right\vert \exp(-\beta H)\left\vert \overrightarrow{r}%
^{,}\right\rangle =%
{\displaystyle\sum_{n}}
\varphi_{n}^{\ast}(\overrightarrow{r})\varphi_{n}(\overrightarrow{r}^{,}%
)\exp(-\beta\varepsilon_{n}) \tag{3}%
\end{equation}
can be obtained through the Laplace transform result%
\begin{equation}
C(\overrightarrow{r},\overrightarrow{r}^{,};\beta)=\beta\int_{0}^{\infty
}d\lambda e^{-\beta\lambda}\rho(\overrightarrow{r},\overrightarrow{r}^{,})
\tag{4}%
\end{equation}
It should be noted that, in quantum statistics and thermodynamics, $\beta$ is
an inverse of temperature: $\beta=1/k_{B}T$ with $k_{B}$ the Boltzmann
constant; but if we now replace $\beta$ in equation (3) by $\beta\rightarrow
it/\hbar$, the resulting propagator $K(\overrightarrow{r},\overrightarrow
{r}^{,};t)$ describes the propagation of the single particle from
$\overrightarrow{r}^{,}\rightarrow$ $\overrightarrow{r}$ in time $t$. However
in the subsequent analysis $\beta$ is to be viewed as a complex parameter. The
interest in the Bloch propagator is that it contains all quantum mechanical
informations \cite{March1959}, \cite{Brack_Bhaduri2003}, \cite{Shea}, from
which the density matrix $\rho(\overrightarrow{r},\overrightarrow{r}^{,})$ in
Eq. (2) may be obtained by suitable inverse Laplace transform, that is \
\begin{equation}
\rho(\overrightarrow{r},\overrightarrow{r}^{,})=\frac{1}{2\pi i}%
{\displaystyle\int\limits_{c-i\infty}^{c+i\infty}}
e^{\beta\lambda}\frac{C(\overrightarrow{r},\overrightarrow{r}^{,},\beta
)}{\beta}d\beta\text{, \ \ \ \ \ }c>0 \tag{5}%
\end{equation}
Note that, to carry out the complex integration in equation (5), the parameter
$\beta$, as stated before, is considered as a complex mathematical variable .

Putting $\overrightarrow{r}^{,}=\overrightarrow{r}$ in equation (4), we
evidently get
\begin{equation}
C(\overrightarrow{r};\beta)=\beta\int_{0}^{\infty}d\lambda e^{-\beta\lambda
}\rho(\overrightarrow{r}) \tag{6}%
\end{equation}
where $C(\overrightarrow{r};\beta)$ denotes the diagonal elements of the Bloch
propagator, called also the Slater sum, and $\rho(\overrightarrow{r})$ is the
particle density.

With the above analysis, all the results are formally exact. In the next
section we shall use these results within the framework of gradient expansion.
Notice that for the case of constant effective mass hamiltonians, the local
version of equation (5) has been used to calculate the density from the Bloch
propagator \cite{Holas}. For Hamiltonians with position dependent mass, we
invert the procedure and use equation (6) to get the Bloch propagator since,
as stated before, the gradient expansions of the density are known.

\section{Semiclassical expansion of the Slater sum for spatially varying
effective mass hamiltonians in dimensions d=1,2,3,4.}

In this section, explicit $\hbar$ expansions will be presented for the Slater
sum through the use of equation (6). For that, we directly use the $\hbar$
expansions of the particle density $\rho(\overrightarrow{r})$ derived in
\cite{Berk_bench2005}. In one spatial dimension it is given, up to order
$\hbar^{2}$, by [see equation (A5) of reference \cite{Berk_bench2005}]%

\begin{align}
\rho_{d=1}\left(  x\right)   &  =\frac{1}{\pi}\sqrt{\frac{2m_{0}}{\hbar^{2}f}%
}\left(  \lambda-V\right)  ^{+1/2}\theta\left(  \lambda-V\right)  +\nonumber\\
\sqrt{\frac{\hbar^{2}f}{2m_{0}}} &  \left\{  \left[  \frac{1}{32\pi}\frac
{1}{f^{2}}\left(  \frac{df}{dx}\right)  ^{2}\left(  \lambda-V\right)
^{-1/2}+\frac{1}{48\pi}\left(  2\frac{d^{2}V}{dx^{2}}+\frac{1}{f}\frac{df}%
{dx}\frac{dV}{dx}\right)  \left(  \lambda-V\right)  ^{-3/2}+\right.  \right.
\nonumber\\
&  \left.  \frac{1}{32\pi}\left(  \frac{dV}{dx}\right)  ^{2}\left(
\lambda-V\right)  ^{-5/2}\right]  \theta\left(  \lambda-V\right)  -\nonumber\\
&  \left[  \frac{1}{24\pi}\left(  2\frac{d^{2}V}{dx^{2}}+\frac{1}{f}\frac
{df}{dx}\frac{dV}{dx}\right)  \left(  \lambda-V\right)  ^{-1/2}+\frac{1}%
{24\pi}\left(  \frac{dV}{dx}\right)  ^{2}\left(  \lambda-V\right)
^{-3/2}\right]  \delta\left(  \lambda-V\right)  +\nonumber\\
&  \left.  \frac{1}{24\pi}\left(  \frac{dV}{dx}\right)  ^{2}\left(
\lambda-V\right)  ^{-1/2}\frac{\partial\delta\left(  \lambda-V\right)
}{\partial\lambda}\right\}  \tag{7}%
\end{align}
Here the potential $V(\overrightarrow{r})$ is related to the one-body
potential $U(\overrightarrow{r})$ in Eq.$(1)$ by $V(\overrightarrow
{r})=U(\overrightarrow{r})+\frac{\hbar^{2}}{8m_{0}}\overrightarrow{\nabla}%
^{2}f(\overrightarrow{r})$ \cite{Berk_bench2005} and $\delta$ is the Dirac
distribution. Notice that here we do not include the spin degeneracy (factor
of two for spin half particles) in the expression of the particle density in
(7) as was done in \cite{Berk_bench2005}. Next, it is easy to verify that
Eq.$\left(  7\right)  $ may be simplified to%
\begin{align}
\rho_{d=1}\left(  x\right)   &  =\frac{1}{\pi}\sqrt{\frac{2m_{0}}{\hbar^{2}f}%
}\left(  \lambda-V\right)  ^{+1/2}\theta\left(  \lambda-V\right)  +\nonumber\\
&  \sqrt{\frac{\hbar^{2}f}{2m_{0}}}\left\{  \frac{1}{16\pi}\frac{1}{f^{2}%
}\left(  \frac{df}{dx}\right)  ^{2}\left[  \frac{\partial\left(
\lambda-V\right)  ^{+1/2}\theta\left(  \lambda-V\right)  }{\partial\lambda
}\right]  \text{ }-\right.  \nonumber\\
&  \frac{1}{24\pi}\left(  2\frac{d^{2}V}{dx^{2}}+\frac{1}{f}\frac{df}{dx}%
\frac{dV}{dx}\right)  \left[  \frac{\partial\left(  \lambda-V\right)
^{-1/2}\theta\left(  \lambda-V\right)  }{\partial\lambda}\right]  +\nonumber\\
&  \left.  \frac{1}{24\pi}\left(  \frac{dV}{dx}\right)  ^{2}\left[
\frac{\partial^{2}\left(  \lambda-V\right)  ^{-1/2}\theta\left(
\lambda-V\right)  }{\partial\lambda^{2}}\right]  \right\}  \tag{8}%
\end{align}
which can be alternatively rewritten as
\begin{align}
\rho_{d=1}\left(  x\right)   &  =\frac{1}{\pi}\sqrt{\frac{2m_{0}}{\hbar^{2}f}%
}\left(  \lambda-V\right)  ^{1/2}\theta\left(  \lambda-V\right)  +\nonumber\\
&  \sqrt{\frac{\hbar^{2}f}{2m_{0}}}\left\{  \frac{1}{16\pi}\frac{1}{f^{2}%
}\left(  \frac{df}{dx}\right)  ^{2}\left[  \frac{\partial\left(
\lambda-V\right)  ^{+1/2}\theta\left(  \lambda-V\right)  }{\partial\lambda
}\right]  \text{ }-\right.  \nonumber\\
&  \frac{1}{12\pi}\left(  2\frac{d^{2}V}{dx^{2}}+\frac{1}{f}\frac{df}{dx}%
\frac{dV}{dx}\right)  \left[  \frac{\partial^{2}\left(  \lambda-V\right)
^{+1/2}\theta\left(  \lambda-V\right)  }{\partial\lambda^{2}}\right]
+\nonumber\\
&  \left.  \frac{1}{12\pi}\left(  \frac{dV}{dx}\right)  ^{2}\left[
\frac{\partial^{3}\left(  \lambda-V\right)  ^{+1/2}\theta\left(
\lambda-V\right)  }{\partial\lambda^{3}}\right]  \right\}  \tag{9}%
\end{align}
We put $\rho_{d=1}\left(  x\right)  $ in the form given by Eq.$\left(
9\right)  $ to show that, for the case of a constant effective mass, our
expression for the density reduces exactly to the one given by equation (12)
of \cite{salasnich} in one spatial dimension. It is interesting to note that,
the latter density was obtained through a different semiclassical method
namely the Kirzhnits expansion.

However, to obtain the local Bloch density, we use Eq.$(8)$ rather than $(9)$
and substitute it into Eq.$\left(  6\right)  $. Then we use the following
property of Laplace transforms \cite{Brack_Bhaduri2003}, \cite{Abraw}
\begin{equation}
\int_{0}^{\infty}d\lambda e^{-\beta\lambda}\left[  \left(  \lambda
-V(\overrightarrow{r})\right)  ^{\nu}\theta\left(  \lambda-V(\overrightarrow
{r})\right)  \right]  =\frac{\Gamma\left(  \nu+1\right)  }{\beta^{\nu+1}%
}e^{-\beta V(\overrightarrow{r})} \tag{10}%
\end{equation}
which we applied to each term of the expansion, we then obtain
\begin{align}
C_{d=1}(\overrightarrow{r};\beta)  &  =\left(  \frac{m_{0}}{2\pi\hbar
^{2}f\beta}\right)  ^{1/2}e^{-\beta V(\overrightarrow{r})}\left[
1+\frac{\hbar^{2}f}{24m_{0}}\right.  \left\{  \frac{3}{4}\left(  \frac{1}%
{f}\frac{df}{dx}\right)  ^{2}\beta+\right. \nonumber\\
&  \left.  \left.  \left[  \left(  -\frac{1}{f}\frac{df}{dx}.\frac{dV}%
{dx}\right)  -2\left(  \frac{d^{2}V}{dx^{2}}\right)  \right]  \beta
^{2}+\left(  \frac{dV}{dx}\right)  ^{2}\beta^{3}\right\}  \right]  \tag{11}%
\end{align}
For the two dimensional case, we write down the corresponding $\hbar$
expansions of the density $\left[  \text{given by Eq.}(28)\text{ of
\cite{Berk_bench2005}}\right]  $ as%
\begin{align}
\rho_{d=2}(\overrightarrow{r})  &  =\frac{m_{0}}{2\pi\hbar^{2}f}%
(\lambda-V)\;\Theta(\lambda-V)\;+\frac{1}{48\pi}\left[  \left(  \frac
{\overrightarrow{\nabla}f}{f}\right)  ^{2}-\left(  \frac{\overrightarrow
{\nabla}^{2}f}{f}\right)  \right]  \Theta(\lambda-V)\nonumber\\
&  -\frac{1}{24\pi}(\overrightarrow{\nabla}^{2}V)\;\delta(\lambda-V)+\frac
{1}{48\pi}\left(  \overrightarrow{\nabla}V\right)  ^{2}\frac{\partial
\delta(\lambda-V)}{\partial\lambda} \tag{12}%
\end{align}
Plugging Eq.$(12)$ into $(6)$ and Laplace transforming\textbf{ (}%
Eq\textbf{.}$\mathbf{(}10\mathbf{)}$\textbf{)}, we find for the local Bloch
propagator
\begin{equation}
C_{d=2}(\overrightarrow{r};\beta)=\left(  \frac{m_{0}}{2\pi\hbar^{2}f\beta
}\right)  e^{-\beta V(\overrightarrow{r})}\left[  1+\frac{\hbar^{2}f}{24m_{0}%
}\left\{  \left[  \left(  \frac{\overrightarrow{\nabla}f}{f}\right)
^{2}-\left(  \frac{\overrightarrow{\nabla}^{2}f}{f}\right)  \right]
\beta-2\left(  \overrightarrow{\nabla}^{2}V\right)  \beta^{2}+(\overrightarrow
{\nabla}V)^{2}\beta^{3}\right\}  \right]  \tag{13}%
\end{equation}
For the $d=3$ case, the expression of the density, up to order $\hbar^{2},$
[see for instance \cite{bencheikh2006} and references cited therein] is given
by%
\begin{align}
\rho_{d=3}(\overrightarrow{r})  &  =\frac{1}{6\pi^{2}\hbar^{3}}\left(
\frac{2m_{0}}{f}\right)  ^{3/2}\left(  \lambda-V\right)  ^{3/2}\Theta
(\lambda-V)+\nonumber\\
&  \frac{1}{24\pi^{2}\hbar}\sqrt{\frac{m_{0}}{2f}}\left\{  \left[  \frac{7}%
{4}\left(  \frac{\overrightarrow{\nabla}f}{f}\right)  ^{2}-2\left(
\frac{\overrightarrow{\nabla}^{2}f}{f}\right)  \right]  \left(  \lambda
-V\right)  ^{1/2}\Theta(\lambda-V)+\right. \nonumber\\
&  \left[  \left(  \frac{\overrightarrow{\nabla}f.\overrightarrow{\nabla}V}%
{f}\right)  -2(\overrightarrow{\nabla}^{2}V)\;\right]  \left(  \lambda
-V\right)  ^{-1/2}\Theta(\lambda-V)\left.  -\frac{1}{4}\left(  \overrightarrow
{\nabla}V\right)  ^{2}\left(  \lambda-V\right)  ^{-3/2}\Theta(\lambda
-V)\right\}  \tag{14}%
\end{align}
Substituting Eq.$(14)$ into Eq.$(6)$, and using Eq\textbf{.}$\mathbf{(}%
10\mathbf{)}$, we get
\begin{align}
C_{d=3}(\overrightarrow{r};\beta)  &  =\left(  \frac{m_{0}}{2\pi\hbar
^{2}f\beta}\right)  ^{3/2}e^{-\beta V(\overrightarrow{r})}\left[
1+\frac{\hbar^{2}f}{24m_{0}}\left\{  \left[  \frac{7}{4}\text{ }\left(
\frac{\overrightarrow{\nabla}f}{f}\right)  ^{2}-2\left(  \frac{\overrightarrow
{\nabla}^{2}f}{f}\right)  \right]  \beta+\right.  \right. \nonumber\\
&  \left.  \left.  \left[  \left(  \frac{\overrightarrow{\nabla}%
f.\overrightarrow{\nabla}V}{f}\right)  -2\left(  \overrightarrow{\nabla}%
^{2}V\right)  \right]  \beta^{2}+(\overrightarrow{\nabla}V)^{2}\beta
^{3}\right\}  \right]  \tag{15}%
\end{align}
It is interesting to observe that the results given respectively in
Eqs\textbf{.}$(11)$, $(13)$ and $(15)$ can be written down in terms of the
dimensionality $d$ \ of the space as follows
\begin{align}
C_{d}(\overrightarrow{r};\beta)  &  =\left(  \frac{m_{0}}{2\pi\hbar^{2}f\beta
}\right)  ^{d/2}e^{-\beta V(\overrightarrow{r})}\times\left[  1+\frac
{\hbar^{2}f}{24m_{0}}\left\{  \left[  \frac{\left(  d-1\right)  ^{2}+3}%
{4}\left(  \frac{\overrightarrow{\nabla}f}{f}\right)  ^{2}+\left(  1-d\right)
\left(  \frac{\overrightarrow{\nabla}^{2}f}{f}\right)  \right]  \beta\text{
}+\right.  \right. \nonumber\\
&  \left.  \left.  \left[  \left(  d-2\right)  \left(  \frac{\overrightarrow
{\nabla}f.\overrightarrow{\nabla}V}{f}\right)  -2\left(  \overrightarrow
{\nabla}^{2}V\right)  \right]  \beta^{2}+(\overrightarrow{\nabla}V)^{2}%
\beta^{3}\right\}  \right]  \tag{16}%
\end{align}
We have looked whether the above equation is valid for higher dimensions or at
least for $d=4.$ For that, we have make use of the semiclassical approach in
Ref. \cite{Grammaticos_79} to write down the density, up to order $\hbar^{2}$,
in $d=4$ dimension. We have obtained
\begin{align}
\rho_{d=4}(\overrightarrow{r})=\frac{m_{0}^{2}}{8\pi^{2}\hbar^{4}f^{2}}\left(
\lambda-V\right)  ^{2}\theta\left(  \lambda-V\right)   &  +\frac{m_{0}}%
{32\pi^{2}\hbar^{2}f}\left[  \left(  \frac{\overrightarrow{\nabla}f}%
{f}\right)  ^{2}-\left(  \frac{\overrightarrow{\nabla}^{2}f}{f}\right)
\right]  \left(  \lambda-V\right)  \theta\left(  \lambda-V\right)
+\nonumber\\
\frac{m_{0}}{48\pi^{2}\hbar^{2}f}\left(  \frac{\overrightarrow{\nabla
}f.\overrightarrow{\nabla}V}{f}\right)  \theta\left(  \lambda-V\right)   &
-\frac{m_{0}}{48\pi^{2}\hbar^{2}f}\left(  \overrightarrow{\nabla}^{2}V\right)
\theta\left(  \lambda-V\right)  +\frac{m_{0}}{96\pi^{2}\hbar^{2}f}\left(
\overrightarrow{\nabla}V\right)  ^{2}\delta\left(  \lambda-V\right)  \tag{17}%
\end{align}
Here $\overrightarrow{\nabla}$ stands for the gradient in four dimensions.
Having the density, we follow the same derivations as done for $d=1,2,3$ to
obtain the corresponding expression of the Slater sum and we found%
\begin{align}
C_{d=4}(\overrightarrow{r};\beta)  &  =\left(  \frac{m_{0}}{2\pi\hbar
^{2}f\beta}\right)  ^{2}e^{-\beta V(\overrightarrow{r})}\times\nonumber\\
&  \left[  1+\frac{\hbar^{2}}{24m_{0}}\left\{  3\left[  \left(  \frac
{\overrightarrow{\nabla}f}{f}\right)  ^{2}\text{ }-\left(  \frac
{\overrightarrow{\nabla}^{2}f}{f}\right)  \right]  \beta+\left[  2\left(
\frac{\overrightarrow{\nabla}f.\overrightarrow{\nabla}V}{f}\right)  -2\left(
\overrightarrow{\nabla}^{2}V\right)  \right]  \beta^{2}+(\overrightarrow
{\nabla}V)^{2}\beta^{3}\right\}  \right]  \tag{18}%
\end{align}
It is easy to check that, setting $d=4$ in Eq.$(16)$, one recovers the result
in Eq.$(18)$. Hence our expression given in Eq.$(16)$ holds true for
$d=1,2,3,4$. Notice that, in Eq.$(16)$, unlike the position dependent mass
terms, the remaining terms involving gradients of the potential do not depend
on the dimension $d$ of the considered space.

Note that the Slater sum in Eq.$(16)$ is expressed in terms of
$V(\overrightarrow{r})=U(\overrightarrow{r})+\frac{\hbar^{2}}{8m_{0}%
}\overrightarrow{\nabla}^{2}f(\overrightarrow{r})$. For practical use, it is
interesting to re-express it in terms of the original one body potential
$U(\overrightarrow{r})$. Upon substitution, equation (16) becomes
\begin{align}
C_{d}(\overrightarrow{r};\beta)  &  =\left(  \frac{m_{0}}{2\pi\hbar^{2}f\beta
}\right)  ^{d/2}e^{-\beta U(\overrightarrow{r})}\left[  1+\frac{\hbar^{2}%
f}{24m_{0}}\left\{  \left[  \frac{\left(  d-1\right)  ^{2}+3}{4}\left(
\frac{\overrightarrow{\nabla}f}{f}\right)  ^{2}+\left(  1-d\right)  \left(
\frac{\overrightarrow{\nabla}^{2}f}{f}\right)  \right]  \beta\text{ }+\right.
\right. \nonumber\\
&  \left.  \left.  \left[  \left(  d-2\right)  \left(  \frac{\overrightarrow
{\nabla}f.\overrightarrow{\nabla}V}{f}\right)  -2\left(  \overrightarrow
{\nabla}^{2}V\right)  \right]  \beta^{2}+(\overrightarrow{\nabla}V)^{2}%
\beta^{3}\right\}  \right]  e^{-\beta\frac{\hbar^{2}}{8m_{0}}\overrightarrow
{\nabla}^{2}f(\overrightarrow{r})}\nonumber\\
&  =\left(  \frac{m_{0}}{2\pi\hbar^{2}f\beta}\right)  ^{d/2}e^{-\beta
U(\overrightarrow{r})}\left[  1+\frac{\hbar^{2}f}{24m_{0}}\left\{  \left[
\frac{\left(  d-1\right)  ^{2}+3}{4}\left(  \frac{\overrightarrow{\nabla}f}%
{f}\right)  ^{2}+\left(  1-d\right)  \left(  \frac{\overrightarrow{\nabla}%
^{2}f}{f}\right)  \right]  \beta\text{ }+\right.  \right. \nonumber\\
&  \left.  \left.  \left[  \left(  d-2\right)  \left(  \frac{\overrightarrow
{\nabla}f.\overrightarrow{\nabla}V}{f}\right)  -2\left(  \overrightarrow
{\nabla}^{2}V\right)  \right]  \beta^{2}+(\overrightarrow{\nabla}V)^{2}%
\beta^{3}\right\}  \right]  \left(  1-\beta\frac{\hbar^{2}}{8m_{0}%
}\overrightarrow{\nabla}^{2}f\right)  \tag{19}%
\end{align}
In getting the second form, use has been made of the Taylor expansion up to
order\textbf{ }$\hbar^{2}$ of exp$(-\beta\frac{\hbar^{2}}{8m_{0}%
}\overrightarrow{\nabla}^{2}f)\approx1-$ $\beta\frac{\hbar^{2}}{8m_{0}%
}\overrightarrow{\nabla}^{2}f$. Since the terms involving $\overrightarrow
{\nabla}V$ and $\overrightarrow{\nabla}^{2}V$ are of order $\hbar^{2}$, we
need only replace them in Eq.$(19)$ by their leading terms, i.e;
$\overrightarrow{\nabla}V$ $=\overrightarrow{\nabla}U+O(\hbar^{2})$ and
$\overrightarrow{\nabla}^{2}V=$ $\overrightarrow{\nabla}^{2}U+O(\hbar^{2})$ .
This, after simple rearrangements up to the order of $\hbar^{2}$, leads
finally to
\begin{align}
C_{d}(\overrightarrow{r};\beta)  &  =\left(  \frac{m_{0}}{2\pi\hbar^{2}f\beta
}\right)  ^{d/2}e^{-\beta U}\left[  1+\frac{\hbar^{2}f}{24m_{0}}\left\{
\left[  \frac{\left(  d-1\right)  ^{2}+3}{4}\left(  \frac{\overrightarrow
{\nabla}f}{f}\right)  ^{2}-\left(  d+2\right)  \left(  \frac{\overrightarrow
{\nabla}^{2}f}{f}\right)  \right]  \beta\text{ }+\right.  \right. \nonumber\\
&  \left.  \left.  \left[  \left(  d-2\right)  \left(  \frac{\overrightarrow
{\nabla}f.\overrightarrow{\nabla}U}{f}\right)  -2\left(  \overrightarrow
{\nabla}^{2}U\right)  \right]  \beta^{2}+(\overrightarrow{\nabla}U)^{2}%
\beta^{3}\right\}  \right]  \tag{20}%
\end{align}
Recall that $f(\overrightarrow{r})=m_{0}/m^{\ast}(\overrightarrow{r})$. As can
be seen the above expression receives explicit contributions from the
spatially varying effective mass $m^{\ast}(\overrightarrow{r})$ not only at
zero order, through the term proportional to $\ f^{-d/2}$ and also from terms
of order $\hbar^{2}$ proportional to $\overrightarrow{\nabla}f$ and
$\overrightarrow{\nabla}^{2}f$. The above equation is the main result of the
present study. For $d=3$, our expression reduces to the one obtained long time
ago in the context of nuclear physics \cite{brack_76}.

\section{Numerical Example}

In this section, we want to numerically test the importance of the position
dependent effective mass terms in the derived Slater sum Eq.$(20)$. For that,
we need as an input, a given effective mass $m^{\ast}$and a potential
$U$.\textbf{ }Without loss of generality\textbf{ }we focus on one-dimensional
systems with mass distribution $m^{\ast}(x)$ and we choose $U(x)$ so that
Eq.$(1)$ possesses an exact analytical solution. In Ref. \cite{Alhaidari_2002}%
, Alhaidari solved \ exactly\textbf{ }Eq\textbf{.}$\mathbf{(}1\mathbf{)}$, for
a large class of potentials, by means of an elegant method called point
canonical transformations PCT. Let us briefly recall this technique. Under the
following PCT, $y=\int\left(  f(x)\right)  ^{-\frac{1}{2}}dx$ and $\varphi
_{n}(x)=\left(  f(x)\right)  ^{-\frac{1}{4}}\psi_{n}(y)$, the Schr\"{o}dinger
equation $(1)$ with spatially mass distribution $m^{\ast}(x)=m_{0}/f(x)$ and
potential $U(x)$, is mapped to a Schr\"{o}dinger equation with a constant mass
$m_{0}$ so that $\left[  -\frac{\hbar^{2}}{2m_{0}}\frac{d^{2}}{dy^{2}%
}+\widetilde{U}(y)\right]  \psi_{n}(y)=E_{n}\psi_{n}(y)$, with
$U(x)=\widetilde{U}(y)+\frac{\hbar^{2}}{8m(x)}\left[  \frac{1}{m(x)}%
\frac{d^{2}m(x)}{dx^{2}}-\frac{7}{4m^{2}(x)}\left(  \frac{dm(x)}{dx}\right)
^{2}\right]  $ and $E_{n}=\epsilon_{n}$. Taking for the constant mass problem,
the harmonic oscillator potential$\widetilde{\text{ }U}(y)=m_{0}\omega
^{2}y^{2}/2$, Alhaidari obtained the exact solutions for a given $m^{\ast}%
(x)$. Let us now take the specific mass distribution used in
\cite{Alhaidari_2002}.
\begin{equation}
m(x)=m_{0}\left(  \frac{\gamma+x^{2}}{1+x^{2}}\right)  ^{2}\text{
\ \ \ \ \ \ \ \ }m(\pm\infty)=m_{0}\text{ \ \ \ } \tag{21}%
\end{equation}
from which we get $f(x)=\left(  \left(  1+x^{2}\right)  /(\gamma
+x^{2})\right)  ^{2}$ with $\gamma$ being a real constant parameter. One then
obtains $\ y=x+(\gamma-1)\arctan(x)$. Note that when $\gamma=1.0$ , Eq.$(21)$
gives a constant effective mass $m(x)=m_{0}$. In terms of $\ f(x)$ and its
derivatives, the above potential $U(x)$ reads then
\begin{equation}
U(x)=\frac{m_{0}\omega^{2}}{2}\left[  x+(\gamma-1)\mathrm{\arctan}(x)\right]
^{2}+\frac{\hbar^{2}}{8m_{0}}\left[  -\frac{d^{2}f(x)}{dx^{2}}+\frac{1}%
{4f(x)}\left(  \frac{df(x)}{dx}\right)  ^{2}\right]  \tag{22}%
\end{equation}

In Figure 1, we plot the ratio $f(x)=m_{0}/m(x)$ from\ Eq.$(21)$ as a function
of the spatial coordinate$\ x$ for $\gamma$ values equal\textbf{
}to$\ 0.6,0.8$ and $1.0$. Setting the parameters $\omega$ and $\beta$ to $1$,
we plot in Figure 2, the quantity $C_{d=1}(x;\beta)$ of Eq.$(20)$ as a
function of $\ x$ for $\gamma\ \mathbf{=}\ 0.6,0.8$ and$\ 1.0$. Here we
consider a particle of unit free mass $\left(  m_{0}=1\right)  $, and the
Planck constant is set to $1$. One can see in this figure that the presence of
spatially varying effective mass may lead to important deviations locally with
respect to the constant mass case. We also display in Figure 3. the second
order $\hbar^{2}$ term of the Bloch density which (using Eq.$(20)$) is given
by $\ \delta C_{d=1}(x;\beta)=\frac{\hbar^{2}f}{24m_{0}}\left(  \frac{m_{0}%
}{2\pi\hbar^{2}f\beta}\right)  ^{1/2}e^{-\beta U}\left\{  \left[  \frac{3}%
{4}\left(  \frac{1}{f}\frac{df}{dx}\right)  ^{2}-\frac{3}{f}\frac{d^{2}%
f}{dx^{2}}\right]  \beta+\left[  \left(  -\frac{1}{f}\frac{df}{dx}.\frac
{dU}{dx}\right)  -2\left(  \frac{d^{2}U}{dx^{2}}\right)  \right]  \beta
^{2}+\left(  \frac{dU}{dx}\right)  ^{2}\beta^{3}\right\}  $. Deviations from
the constant effective mass case are clearly exhibited at this order.

This being done, it is interesting to observe that, for the above exactly
solvable model in $1d$, one can derive an exact analytical expression for the
Slater sum. Due to the simple relationship between\ the $\varphi_{n}%
^{^{\prime}}s$ and the $\psi_{n}^{^{\prime}}s$ given by $\varphi
_{n}(x)=\left(  f(x)\right)  ^{-\frac{1}{4}}\psi_{n}(y),\ $and since the
position dependent mass problem and the constant mass ones have the same
energy spectrum, we can immediately write\textbf{ }%
\begin{equation}
C(x;\beta)=\frac{1}{\sqrt{f(x)}}\widetilde{C}\left(  y;\beta\right)  \tag{23}%
\end{equation}
where\textbf{ }$\widetilde{C}\left(  y;\beta\right)  =\sum_{n}\left\vert
\psi_{n}(y)\right\vert ^{2}e^{-\beta\epsilon_{n}}$\textbf{ }is the Slater sum
of the problem with constant mass and with potential$\widetilde{\text{ }%
U}(y)=m_{0}\omega^{2}y^{2}/2$\textbf{. }An exact expression of $\widetilde
{C}\left(  y;\beta\right)  $ is known, it reads \cite{BrackPRL2000}\textbf{ }%

\begin{equation}
\widetilde{C}\left(  y;\beta\right)  =\left(  \sqrt{\frac{m_{0}\omega}%
{2\pi\hbar\sinh\left(  \beta\hbar\omega\right)  }}\right)  \exp\left[  \left(
-\frac{m_{0}\omega}{\hbar}\tanh\left(  \frac{\beta\hbar\omega}{2}\right)
\right)  y^{2}\right]  \tag{24}%
\end{equation}
Substituting Eq.$(24)$ into $(23)$, and remembering that, $y=x+(\gamma
-1)\arctan(x)$ we find the exact result%
\begin{equation}
C(x;\beta)=\left(  \sqrt{\frac{m_{0}\omega}{2\pi\hbar f(x)\sinh\left(
\beta\hbar\omega\right)  }}\right)  \exp\left[  \left(  -\frac{m_{0}\omega
}{\hbar}\tanh\left(  \frac{\beta\hbar\omega}{2}\right)  \right)
(x+(\gamma-1)\arctan(x))^{2}\right]  \tag{25}%
\end{equation}
A plot of the above density, is given in figure 4 for different values of
$\gamma$.


\section{Conclusion}

In the present study we have derived the gradient expansion up to order
$\hbar^{2}$, of the Slater sum for hamiltonians with position dependent mass.
Our main result is summarized by Eq\textbf{.}$(20)$ valid for spatial
dimensions $d=1,2,3,4.$ This result is valid for arbitrary\ one-body potential
$U(\overrightarrow{r})$ and spatially varying effective mass $m^{\ast
}(\overrightarrow{r})$.

We have explicitly shown that, for $d=1$ and constant effective mass the
expressions of the semiclassical densities derived here within the algebraic
method is identical to\ the Kirzhnits expansion up to order $\hbar^{2}$. This
would be certainly true for higher dimensions. The results we have obtained
would constitute useful approximation to the exact calculation of propagator
or Green functions \cite{alhaidari}, \cite{chetouani2009} for hamiltonians
with spatially varying mass.\bigskip\ 

Very recently a gradient expansion (up to second order) of density matrix
$\rho(\overrightarrow{r},\overrightarrow{r}^{\prime})$ has been obtained in
two dimensions \cite{Rasanen_2012} and one may use equation $(10)$ to get the
full Bloch propagator $C(\overrightarrow{r},\overrightarrow{r}^{,};\beta)$.
For the $3d$ case, the gradient expansion was derived long time ago
\cite{Dreizer}. An interesting extension is to generalize these results to
include a position dependent mass in the considered Hamiltonian.

%

\begin{figure}
[ptb]
\begin{center}
\includegraphics[
height=2.7899in,
width=3.998in
]%
{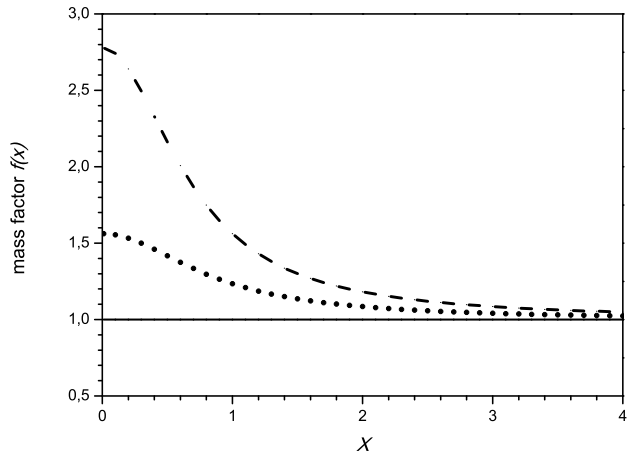}%
\caption{A plot of position dependent effective mass ratio $f(x)=m_{0}%
/m^{\ast}(x)=\left(  \left(  1+x^{2}\right)  /(\gamma+x^{2})\right)  ^{2}$\ .
The solid curve is at value $\gamma=1$, the dot curve is at value $\gamma=0.8$
and the dash curve is at value $\gamma=0.6$ .}%
\end{center}
\end{figure}
%

\begin{figure}
[ptb]
\begin{center}
\includegraphics[
height=2.789in,
width=3.9349in
]%
{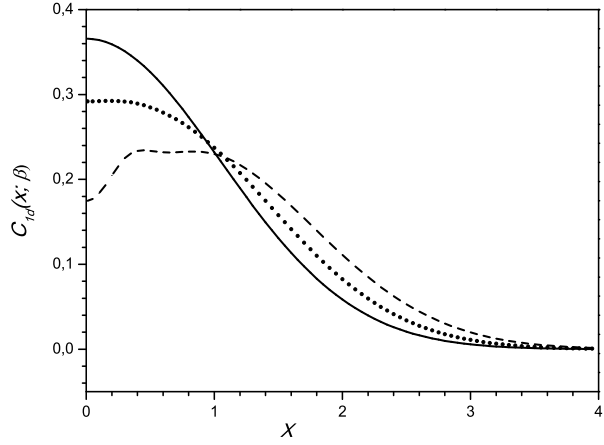}%
\caption{A Plot of the semiclasical Bloch density $C_{d=1}(x;\beta)$ with
$\omega=1,$ $\beta=1$. Solid line correspond to $\gamma=1$ , doted line
$\gamma=0.8$ and dashed line $\ \gamma=0.6$ . \ \ \ \ \ }%
\end{center}
\end{figure}
\begin{figure}
[ptb]
\begin{center}
\includegraphics[
height=2.789in,
width=3.9349in
]%
{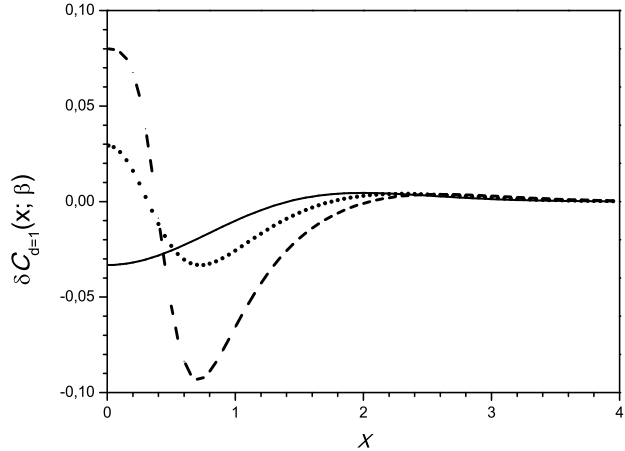}%
\caption{The second order contribution to the Bloch density in figure 2
$\delta C_{d=1}(x;\beta)\ $for values of $\ \gamma=0.6,0.8$ and$\ 1.0$
corresponding respectively to dashed, doted and solid lines.}%
\end{center}
\end{figure}
\begin{figure}
[ptb]
\begin{center}
\includegraphics[
height=2.7899in,
width=3.998in
]%
{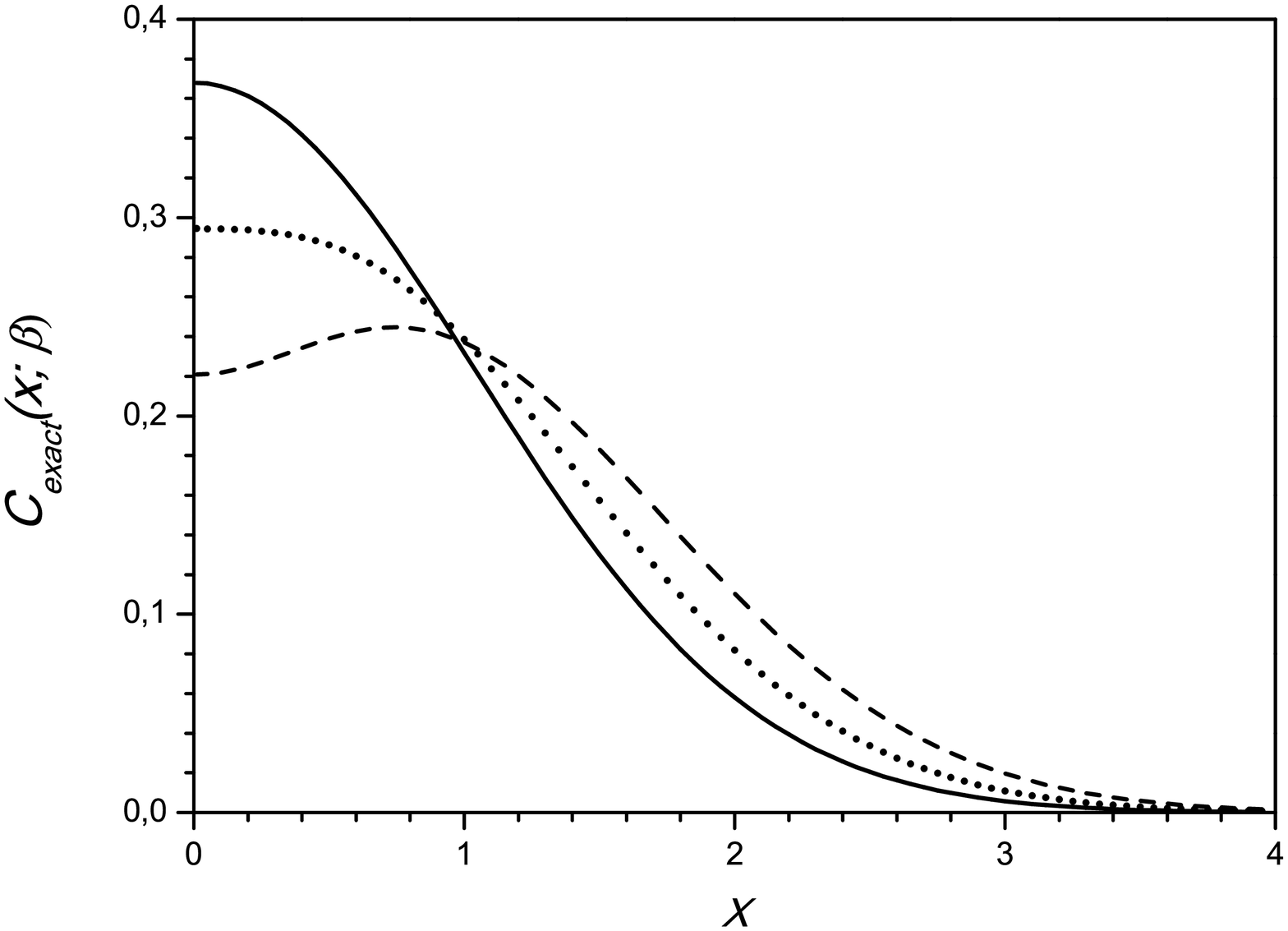}%
\caption{A plot of the exact Slater sum $C_{d=1}(x;\beta)$ [Eq.$(25)$] with
$\omega=1,$ $\beta=1$ for values of $\ \gamma=0.6,0.8$ and$\ 1.0$
corresponding respectively to dashed, doted and solid lines.}%
\end{center}
\end{figure}

\end{document}